\documentclass[12pt,preprint]{aastex}
\usepackage{psfig}
\newcommand{\sso}{$^{-1}$}

\slugcomment{To be published in the Astrophysical Journal, January 2003}

\begin{document}

\title{The Formation of Nuclear Rings in Barred Spiral Galaxies}
\author{Michael W. Regan\altaffilmark{1} \&
	Peter Teuben\altaffilmark{2} 
	}
\altaffiltext{1}{Space Telescope Science Institute, 3700 San Martin Drive,
	Baltimore, MD 21218, mregan@stsci.edu}
\altaffiltext{2}{Department of Astronomy, University of Maryland, 
	College Park, MD 20742, teuben@astro.umd.edu}

\begin{abstract}
Although nuclear rings of gas and star formation are common in
barred spiral galaxies,
current theories of why and how they form do not provide the
level of detail needed to quantify the effect that these rings
can have on the fueling of active galactic nuclei and on the
evolution of their host galaxy.
In this paper we use detailed modeling to
show that existence of nuclear rings is 
directly related to the existence of the orbit family
whose major axis is perpendicular to the major axis of the bar (x$_2$).
We explore a large range of barred galaxy potentials and
for each potential we use a two-dimensional hydrodynamic simulation to
determine whether
and at what radius a nuclear ring forms.
We compare the results of the
hydrodynamic simulations to numerical integrations of 
periodic orbits in a barred
potential and show that the rings only form when 
a minimum amount of x$_2$ orbits exists.
Because the rings migrate inwards with time as they accumulate gas, 
the radius at which a nuclear ring is seen does not give
direct information on the shape of the rotation curve.
We also show that the common assumption that nuclear rings are related to
an inner Lindblad resonance is incorrect.
In fact, we show that there is no \emph{resonance} at the
inner Lindblad resonance in barred galaxies.
We also compare
the predictions of this theory to HST observations and show that
it correctly predicts the observed gas and star formation morphology of nuclear rings.
\end{abstract}
\keywords{ISM: kinematics and dynamics --- galaxies: structure --- 
galaxies: kinematics and dynamics --- galaxies: nuclei --- galaxies: starburst }

\section{Introduction}
Nuclear rings are regions of high star formation rates in the centers of
barred spiral galaxies.
This enhancement in the star formation rate can be high enough that 
these rings contain a majority of the current
star formation in the entire galaxy \citep{GB91}.
These rings affect the evolution of their host galaxy through a
variety of mechanisms.
For example,
models of gas flow in barred galaxies show that 
gas driven toward the nucleus of a barred spiral galaxy can be
interrupted by a nuclear ring \citep{PST95} preventing
the fueling of an active galactic nucleus by the bar.
On the other hand, the formation of new stars in the nuclear ring of the
galaxy can affect the Hubble type of the galaxy.
If enough new stars form from gas brought to the central regions by the
bar, a galaxy could evolve to an earlier Hubble type \citep{NSH96}. 
Therefore, we need a detailed theory of how and why these  rings
form to quantify both the interaction of bar driven gas inflow and nuclear rings
and its effect  
on galaxy evolution and the fueling of active galactic nuclei.

Although the conventional wisdom is that nuclear rings are associated
with the Inner Lindblad Resonance (ILR) of the
large scale bar (see review by \cite{BC96}  and references therein),
the details of this association vary between various theories.
All of these theories explain the formation of the
rings in a qualitative way rather than from first principles.
This lack of detail in the theories is manifested in their inability
to explain
why some galaxies which meet the criteria of the theories for ring formation
have nuclear rings while others that meet the criteria do not have nuclear rings.

The most widely accepted theory for nuclear ring formation
is that the gas in the nuclear ring is trapped either between an
inner ILR (IILR) and an outer ILR (OILR) or at a single ILR
if the potential only has one ILR \citep{Combes96,BC96}.
This theory is based on the claim
that the torque on the gas changes sign at each
resonance \citep{Combes96}.
This implies that
for the case of a single ILR,
gas inside the ILR feels a positive torque and thus gains
angular momentum and moves outward, while
gas between the ILR and corotation receives a negative torque and flows
inward, forming a ring at the ILR.
When there are two ILRs \cite{BC96} state that gas will accumulate at the
IILR.
They invoke two possible reasons for this, either that gas dissipation causes
the gas to flow inward to the IILR or that transient spirals forming between
the ILRs are dominated by leading spirals which cause the gas to flow inward.
Because of the predicted change in sign of the torque on the gas, if this theory is
correct, all galaxies with gas and one or more ILRs should form nuclear rings.

Another theory is that nuclear rings form in the central regions of barred galaxies
because the torque from the large scale bar weakens near the
center of the bar \citep{SBF90}.
The overall potential becomes less bar-like as the influence of the more axisymmetric
bulge begins to dominate the disk potential containing the bar perturbation.
This theory claims that ring-like gas distributions result at the positions
of resonances because at the resonances no net gravitational torques are
applied to the gas.

An alternative explanation to nuclear rings being a feature of the bar and the gravitational potential
is that they are the remnants of a nuclear starburst \citep{KCY93}.
Because the surface density of gas is higher in the central regions of nuclear starbursts
and the star formation rate is proportional to the gas surface density to a power greater than one,
the expectation is that this region would use up its fuel faster, leaving the nuclear
ring as a remnant.

These various theories of ring formation have been explored primarily using
numerical simulations of gas flow in barred galaxies.
The majority of these simulations used a ballistic cloud or sticky particle model of the 
interstellar medium (ISM)
and a fixed gravitational potential \citep{S81,S84,CG85,B94,RS00}. 
This type of modeling of the ISM has the advantage of low numerical viscosity
and relatively high spatial resolution.
The low viscosity
allows the rings to form in the simulations without being quickly driven to the center of the galaxy
by the lost angular momentum due to numerical viscosity (a problem 
which plagued early grid-based methods).
It has the disadvantage of its rather ad hoc modeling of the ISM as
a collection of a low number of collisional particles rather than as a gas.
While these studies investigated all types of rings (outer, inner, and nuclear),
they all associated the formation of nuclear rings with the existence of
an ILR.
Even so, \cite{RS00} found evidence that more than just an ILR is 
required to form a nuclear ring when strong bars in their simulations did not form rings.

The first study to use a grid-based ideal gas hydrodynamic model of the ISM that
was able to form a nuclear ring was by \cite{PST95}.
The problem of numerical viscosity was solved by using higher order solutions to
the hydrodynamic equations, which became practical with faster processors;
the numerical viscosity in this simulation was too low to be measured.
The needed spatial resolution was obtained by
switching to a cylindrical coordinate system, which allowed more resolution than a Cartesian grid
in the nuclear region where the rings form.
Note that subsequent hydrodynamic models with higher resolution 
Cartesian grids have also been able to
create nuclear rings (van Albada 1996, private communications).
From their hydrodynamic simulations, they predicted where nuclear rings would form
and how strong they would be.
First, they claimed that the higher the peak of the $\Omega-\kappa/2$ curve is
above the pattern speed of the bar, the stronger the ring will be.
They also claimed that the ring will form at the radius that corresponds to the
peak of the  $\Omega-\kappa/2$ curve.
Even so, these predictions are based on models of only a few different potentials.
In an additional study that explored more parameter space using the same
code, \cite{Sheth00} found that the ring did not always form
at the location of the peak of the  $\Omega - \kappa/2$ curve but that
the radius of the nuclear ring was proportional to the radius of the outer ILR.

The diversity and inconsistencies of theories, all of which lack a firm quantitative theoretical
foundation for
why and where nuclear rings form, led us
to investigate the formation of nuclear rings in more detail.
One piece of evidence for the formation was already provided by
\cite{A92b} who showed that the existence of offset bar dust lanes is related to the
existence of stable stellar bar orbits whose major axis is perpendicular to the major axis
of the bar (x$_2$ orbits in the nomenclature of \cite{CP80}).
Because nuclear rings in all of the hydrodynamic and sticky particle models
almost always have offset dust lanes,
we would expect there to be
a similar relationship between x$_2$ orbits and nuclear rings.
To show the relationship between the major families of bar orbits we plot a subset of
the orbits in each family in
Figure \ref{x1x2orbits}.
The fact that gas trapped near these x$_2$ orbits cannot co-exist at radii where there is gas
on the orbits elongated parallel to the bar major axis (x$_1$) led us to investigate
the relationship between x$_2$ orbits and nuclear rings.
We find that, rather than the existence of an ILR causing nuclear rings to 
form, the existence of x$_2$ orbits is the true cause of nuclear rings.
Only when a threshold fraction of phase space is occupied by
x$_2$ orbits do rings form, and they initially form
at the radius of the x$_2$ orbit with the largest extent along the bar major axis.

\section{Hydrodynamic Modeling of Ring Formation}

We have chosen to simulate the ISM using a hydrodynamic model because it
provides a better match to kinematic observations of the gas in barred
spiral galaxies than the sticky particle model \citep{RVT97,RSV99}.
Past hydrodynamic studies investigating the gas flow in a barred potential
have either explored a large region of parameter space but with poor
resolution in the nuclear region \citep{A92b} or simulated only a few
potentials with high spatial resolution \citep{PST95}.
To disentangle the various possible characteristics of a barred potential, 
some of which are not orthogonal,
that affect nuclear ring formation requires a large range of
parameter space to be explored.
In addition, relatively high spatial resolution is needed to resolve the
nuclear rings which can have widths
 measured in tens of parsecs.
To investigate this parameter space we ran a series of
two-dimensional high resolution hydrodynamic models using the code of \cite{PST95}.
The only change to the hydrodynamic code was to check the gas surface
density in each cell at every time step, and if it fell below a threshold
($10^{-6}$ M$_{\sun}\ $ pc$^{-2}$), mass was added to bring it back up 
to that threshold.
This change was needed to prevent floating point underflows for some
of the strong bar potentials.
The net mass added to the simulation is not significant, and the actual
minimum gas surface density of the ISM in the central regions of galaxies
is probably much higher.

The potential that was used was the same one as that used by \cite{A92a} and
\cite{PST95}.
The potential is made up of three parts: a disk, a bulge, and a bar.
The disk is a Kuzmin-Toomre disk \citep{K56,T63} with a surface density of
\begin{equation}
\Sigma(r) = (v^{2}_{0}/2\pi G r_{0})(1 + r^2/r^{2}_{0})^{-3/2}
\end{equation}
where v$_0$ = 200 km s\sso\ and r$_0$=14.1 kpc. 
The bulge has a volume density of 
\begin{equation}
\rho(r) = \rho_b(1 + r^2/r^2_b)^{-3/2}
\end{equation}
where $\rho_b$ is the central density of the bulge and r$_b$ is the scale length of the
bulge.
The bar is modeled as a Ferrers ellipsoid along the y axis with a volume density of
\begin{equation}
\rho(g) = \left\{ \begin{array}{ll}
	\rho_0(1 - g^2) \mbox{ for $g < 1$}\\
	0\ \ \ \ \mbox{elsewhere}
       \end{array}
	\right.
\end{equation}	
where $\rho_0$ is the central density of the bar,
$g^2 = y^2/a^2 +x^2/b^2$ and $x$ and $y$ are the 
Cartesian coordinates.
By holding the total mass inside of 10 kpc constant,
the potential can be described by four free parameters: the corotation radius,
the axis ratio of the bar, the quadrupole moment of the bar, and 
the central concentration \citep{PST95}.
To limit the exploration of parameter space to the fast bars that create
offset dust lanes \citep{A92b}, we have set the corotation radius to be 1.2 times the
radius of the bar; for these simulations we have held the bar radius fixed at
5 kpc.

Fixing the corotation radius leaves us with three free parameters which we 
explored through all combinations.
We used five values for the bar axis ratio: 1.5, 2.0, 2.5, 3.0, and 3.5.
For the central concentration we used six different values:
(1.0, 1.5, 2.0, 2.5, 3.0, 3.5) x 10$^{10}$ M$_{\sun}$ kpc$^{-3}$.
For the bar quadrupole moment we used (2.5, 5.0, 7.5, 10, 12.5) x 10$^{10}$
 M$_{\sun}$ kpc$^{2}$.
All combinations of these three parameters yields 150 unique potentials but
12 of these potentials are not physical.
These non-physical potentials result 
for the cases with a bar axis ratio of 1.5 and bar quadrupole moments
 of 10 and 12.5  x 10$^{10}$ M$_{\sun}$ kpc$^{2}$.
These cases require more mass in the bar than there is in the
entire disk. 
Therefore, these 12 cases were not run leaving 138 unique potentials.

All the hydrodynamic simulations were run at what \cite{PST95} termed
high resolution.
At this resolution there are 251 radial zones and 154 angular zones in
one-half of the bar. 
At the inner boundary of the simulation (r=100 pc) the zones are 2x2 pc and at the
end of the bar they are 100x100 pc.
The bar was slowly added into the potential over the initial 100 Myr of the
simulation to minimize transient effects.
In Figures \ref{hydro1}-\ref{hydro5} we show the gas surface densities 
as gray scales
for the inner 5 kpc at a time of 250 Myr.
The reason for showing these surface densities at this relatively 
early time is explained in \S\ref{dentimer} below.

For the case of the bar axis ratio equal to 1.5, we find that
the central morphology looks more like tightly wound spiral arms
rather than a nuclear ring.
This is because the density in the ring shows a large
variation with azimuth.
The density is highest at the end of the bar dust lanes 
along the bar minor axis,
while along the bar major axis the ring becomes very diffuse.

The rings that form when the bar axis ratio is larger than 1.5
have a more
constant density as a function of azimuth.
For these higher axis ratios the galaxy does not always
form a ring.
The overall pattern of the simulations is that nuclear rings are easier to
form when the bar axis ratio is smaller.
In addition, increasing central mass concentration seems to favor 
nuclear ring formation.
In contrast, increasing the bar quadrupole moment seems to inhibit
nuclear ring formation.

\section{Nuclear Rings and Inner Lindblad Resonances}
We can use the hydrodynamic simulations to test some of the
theories for the formation of nuclear rings.
In particular, the results of the simulations allow us to investigate
the relationship between ILRs and nuclear rings.

\subsection{The Time Evolution of the Radii of Nuclear Rings\label{dentimer}}
If, as is the expected understanding, nuclear rings are related to a
resonance, then in our simulations 
the radius of the ring would remain constant with time.
This is because our
potentials do not change with time, keeping the locations of any
resonances constant.
In fact, our simulations show the ring migrating inward with time.
To show this,
we let one of our simulations run 2 Gyr and
at every 10 Myr we take a snapshot of the density in the disk.
We then average the densities at each radius and plot the average density at each
radius as a function of time in Figure \ref{denradtime}.
From the initially smooth density distribution the ring
forms at a radius of around 0.5 - 0.7 kpc.
As the simulation evolves, the radius of the ring decreases until after
about 1 Gyr it is in the range of 0.2 - 0.3 kpc.
This decrease of the radius of the nuclear ring with time has been seen
in other simulations \citep{PST95,RS00}.
\cite{PST95} claimed that the ring evolved inward until it was at
the peak of the $\Omega$-$\kappa$/2 curve but this was based on only one 
simulation.
\cite{RS00} were concerned that this decrease in the ring radius was
a problem with the sticky particle method in a high density environment.
In our simulations all of the rings migrate inward with time showing that
this phenomenon is neither a special case nor a 
numerical simulation problem.

This change of the ring radius with time implies that one 
cannot use the radius of a nuclear ring to infer anything
about the rotation curve.
This means that an observation of the radius of a ring does
not tell you where resonances or other features of the
rotation curve are located.
The migration also provides evidence that the ring is not in an equilibrium state.
Instead something happens with time that forces the ring inward.

The inward migration of the ring is caused by gas flowing down
the dust lane on x$_1$-like streamlines \citep{RVT97}.
This is shown in Figure \ref{dentimecutoff} where we modified the
hydrodynamic simulation to lower the surface density of gas at
all radii greater than 1 kpc to be less than 1$\times 10^{-5} $M$_{\sun}\ $pc$^{-2}$
after 300 Myr.
We waited for 300 Myr to let the ring form, and then, by lowering the
density outside of the nuclear region, we cutoff the inflow of gas.
Figure \ref{dentimecutoff} reveals that the radius of the ring remains constant for 
1.7 Gyr when there is no addition of gas to the ring.
This constant radius with time also rules out
numerical viscosity as the cause of the inward migration.

The inward migration of the ring is the reason why we show the gas
density at 250 Myr in Figures \ref{hydro1}-\ref{hydro5}.
If we had shown the gas density at some later time when the gas flow has
settled down more, the rings would have evolved to smaller radii.
Note that this inward migration of the rings means that the gas flow and
morphology never reach a steady state.

\subsection{ILRs and Nuclear Rings}
If nuclear rings form as the result of an interaction with the ILR then
one would expect a relationship between the existence and 
radii of the ILRs and the existence and radii of nuclear rings.
In Figures \ref{hydrox21}-\ref{hydrox25} we overlay  
on the gas densities from the hydrodynamic simulations
the radii of 
what are commonly referred to as the inner ILR and the 
outer ILR.
We have calculated these radii using the average radius of the ILRs found
from ``rotation curves'' along the bar major and minor axes. 
We find only a weak relationship between the radii and existence of nuclear
rings and the radii and existence of ILRs.

If nuclear rings form due to a resonance at the ILR, then
potentials with ILRs should have nuclear rings.
In fact, we find that ILRs \emph{do} exist for almost all of the potentials that do \emph{not} have
nuclear rings.
Although there are double ILRs in 136 out of the 138 potentials,
we find nuclear rings in only 72 of the potentials.
The existence of ILRs and rings does seem to follow the same general trend 
in that both seem to avoid 
potentials with large bar quadrupole moments and small central mass concentrations.
However, the predicted ILRs exist over a much larger region of parameter 
space than the rings.

When a nuclear ring does form
there is only a weak relationship between the radii of the two
ILRs and the radius of the ring.
Although when rings form they are always between the two ILRs,
the radius of the ring relative to the two ILRs changes in
the different potentials.
There is a general trend that as the bar quadrupole moment increases
the ring radius becomes smaller relative to the ILRs. 
The relative ring radii range from 
the case with the smallest nuclear ring (a/b =3.0, Q$_m$= 10,
$\rho_{cen}$=3.5, run 114), where the nuclear ring radius is only slightly larger
than the IILR,
to elliptical rings which extend out along the bar minor axis to
$\sim$90\% of the OILR radius ($\sim$3 times the IILR radius, a/b =1.5, 
 Q$_m$= 2.5, $\rho_{cen}$=2.0, run 3).
This variation of the ring radius relative to the IILR is not consistent with
the predictions of \cite{BC96} that the nuclear ring forms near the IILR.

\section{X$2$ Orbits and Nuclear Rings}

\subsection{X$_2$ Orbit Extent}

Given the strong relationship between offset bar dust lanes and the
existence of x$_2$ orbits \citep{A92b}, and the fact that these 
dust lanes
intersect nuclear rings at the bar minor axis 
(Figures \ref{hydro1}-\ref{hydro5}),
we decided to investigate the relationship between nuclear rings and x$_2$ orbits.
We find that, unlike ILRs, the radii and existence of x$_2$ orbits are strongly related to the 
radii and existence of nuclear rings.

For each of the 138 potentials we determined the existence and
extent of the periodic x$_2$ orbits. 
Periodic x$_2$ orbits do not exist in all barred galaxies and when they do exist they 
have a limited range of radii over which they extend.
To find the x$_2$ orbit family for each potential 
we began by searching for the x$_2$ orbit with the smallest energy.
An x$_2$ orbit was defined as an orbit with an axis ratio of less than 0.97;
this criterium was found to uniquely identify x$_2$ orbits
in surfaces of section of both x$_1$ and x$_2$ orbits.
If an x$_2$ orbit was found, it would be the one with the smallest radius. 
Using this orbit as a starting point we used the PERORB routine of NEMO \citep{Teuben95}
to find all the periodic orbits in the family.
An example of an x$_2$ orbit family is shown in Figure \ref{x1x2orbits}.

In Figures \ref{hydrox21}-\ref{hydrox25} we overlay on the gas surface
density from the hydrodynamic simulations 
three x$_2$ orbits: the orbit with the largest extent on the bar minor
axis, the orbit with the largest extent on the bar major axis, and
the orbit with the smallest extent on the bar minor axis.
There is an excellent correlation between the x$_2$ orbit with the
largest extent along the bar major axis and the radius of the nuclear
ring.
In all cases when there are no x$_2$ orbits, nuclear rings do not
form.
The overall pattern that emerges from the combination of x$_2$ orbits and
hydrodynamic simulations is that the nuclear ring initially forms at the location
of the x$_2$ orbit with the largest extent along the major axis of the bar.
In some cases where the ring is not circular it extends further along the
minor axis, but it never has the axis ratio of the most elliptical of
the x$_2$ orbits.
Therefore, the nuclear ring is forming at the interface of the x$_1$ and
x$_2$ gas at the largest radius that x$_2$ orbits can exist without crossing
other x$_2$ orbits.

In weak bars, \cite{vAS82} have shown that x$_2$ orbits exist at radii where
 $\Omega_b < \Omega-\kappa/2$ where $\Omega_b$ is the bar pattern speed,
$\Omega$ is the angular frequency of a circular orbit, and $\kappa$ is the
radial epicyclic frequency.
Our simulations also show that this is true for weak bars.
Hereafter, we will refer to the radius where the above equation predicts the smallest radius
x$_2$ orbit as the Inner Epicyclic x$_2$ Radius (IEX2R) and the radius
where the above equation predicts the largest x$_2$ radius as the Outer
Epicyclic x$_2$ Radius (OEX2R).
Note that we prefer this new nomenclature rather than the standard IILR and OILR
due to the arguments in \S\ref{barilr}.
Therefore, the blue circles in
Figures \ref{hydrox21}-\ref{hydrox25} are really the locations
of the IEX2R and the OEX2R rather than the IILR and OILR.
Figures  \ref{hydrox21}-\ref{hydrox25}
show that for cases where the bar axis ratio is not large
(a/b = 1.5 or 2.0), the x$_2$ orbit with the largest extent along the
bar minor axis extends out to the OEX2R.
This approximation breaks down as the bar axis ratio and
quadrupole moment increase.
Also, in many cases with large quadrupole moments and low central
mass concentrations there is a region where $\Omega_b < \Omega-\kappa/2$
but no x$_2$ orbits exist.

There do exist cases at the boundary between no nuclear ring and
clear rings at almost all values of the bar axis ratio where 
there are x$_2$ orbits  but nuclear rings do not form.
In these cases Figures \ref{hydrox21} - \ref{hydrox25} 
provide evidence that there
are relatively few x$_2$ orbits.
This
can been seen by
the small difference between the smallest radius of x$_2$ orbits
and the largest radius.
This implies that not enough gas can exist on these x$_2$ orbits to
prevent the gas on x$_1$ orbits from flowing right through the
x$_2$ region.

To better quantify the minimum amount of x$_2$ orbits needed for a 
nuclear ring to form we have calculated for each potential 
the relative amount of phase 
space that can be occupied by x$_2$ and x$_1$ orbits.
We find
that although x$_2$ orbits must exist for there to be a nuclear ring,
if the available phase space for x$_2$ orbits 
is not large enough nuclear rings do not form.
For each energy where a x$_2$ orbit exists we found the non-closed
x$_2$ orbit that occupied the largest area in the surface of section.
All other x$_2$ orbits lie within this region of the surface of section, while
the x$_1$ orbits occupy the rest of the surface of section.
We then normalized this area by dividing it by the total area of the
surface of section for this energy.
We then summed this
unit-less relative area of x$_2$ orbits over all energies at which x$_2$
orbits exist. 
The resulting sum has the units of the energy step size over which the sum
was taken or 2000 (km s$^{-1}$)$^2$.
The final sums are printed in the upper left of each panel in Figures
\ref{hydrox21} -\ref{hydrox25}.
We see that the threshold for the formation of nuclear rings is
approximately 0.7 to 0.9  (2000 km s$^{-1}$)$^2$ of total phase space volume.

\section{Discussion}

\subsection{Why do Nuclear Rings form?}

Figures \ref{hydrox21}-\ref{hydrox25} show the strong relationship between the x$_2$ orbits
and nuclear rings.
But what is the reason that nuclear rings form?
There are several pieces of evidence that 
the rings form due to the dynamic interaction of gas on x$_1$-like streamlines
and gas on x$_2$-like streamlines.

The closed orbit families shown in Figure \ref{x1x2orbits}
were for stars where the
crossing of x$_1$ and x$_2$ orbits
does not cause collisions due to the low
probability that two stars will actually be 
at the intersection of
their orbits at the same time.
But gas cannot exist on both x$_1$-like streamlines and x$_2$-like streamlines that intersect.
Therefore, when x$_2$ orbits exist, gas can only be on either
the x$_1$-like streamlines or the x$_2$-like streamlines in the region where both are stable.
The simulations show that when enough phase space exists for the x$_2$ orbits, the 
gas will settle on the x$_2$ orbits.
The settling onto x$_2$ orbits is consistent with the empirical rule of
\cite{vAS82} that gas is trapped on the orbit that
is most nearly circular; in this region of barred galaxies this is the x$_2$ orbit.
The most likely reason for this is that at any given radius the x$_2$ orbit 
always has lower energy than the x$_1$ orbit \citep{A92a}.
Even so, clearly this is not a simple equilibrium situation.

The fact that the ring is not in an equilibrium state was shown by Figure
\ref{denradtime} where the radius of the ring decreases with time.
In addition,
the simulations show that gas is accumulating in the ring as time progresses because
gas is flowing down the bar dust lanes along x$_1$-like streamlines \citep{RVT97}. 
It is either the accumulation of this low specific
angular momentum gas in the ring or the dissipation at the contact point of
the inflowing gas and the nuclear ring that causes the radius of the ring to decrease with time.
For example, for the simulation shown in Figure \ref{denradtime}
the angular momentum inside of 1 kpc increases by only 12\% from 300 Myr to 2 Gyr.
But the mass inside of 1 kpc increases by a factor of 3.3.
Also, the requirement that there must be a threshold amount of phase space of x$_2$ orbits 
for a nuclear ring to form
implies that, below this threshold, gas on x$_1$ orbits can prevent the
ring from forming.

Figure \ref{denradtime} also shows that 
the inward migration slows dramaticly when the ring reaches the major axis
of the smallest x$_2$ orbit at around 1 Gyr. 
After that the ring begins to extend over a larger range in radius and seems to oscillate
as the radius of the ring approaches the inner radius of x$_2$ orbits.

Even though the lack of star formation in the models leads to the mass of the ring 
increasing with time, 
the rate at which the model rings accumulate gas ($\sim$1 M$_\sun$ yr\sso) is approximately
the same as the rate of star formation seen in nuclear rings \citep{Maoz01}.
Even so,
the net decrease of the specific angular momentum 
of the gas in nuclear rings of real galaxies would still
cause the radius of rings to decrease with time. 
This implies that the lifetime of the ring is inversely related to rate of mass inflow into the ring.
Therefore, early Hubble type galaxies with low net inflows of gas may have longer lived
rings than later Hubble types due to less gas being available to flow into the ring.

\subsection{The Kinematics of the Gas and Stars in Nuclear Rings}

Because the gas in nuclear rings is being confined there due to the interaction
between gas on x$_1$-like streamlines exterior to the ring and x$_2$-like 
streamlines interior to the ring, it is not in a simple stable orbit.
To show the kinematics of the gas in the nuclear ring, in Figure 
\ref{allfour} we plot the
radial and tangential velocity along the major and minor axes of the
bar for run 100.
The figure shows that along the major axis of the bar, kinematics do not show
any strong transition at the nuclear ring. 
This can also be seen morphologically in many of the simulations where the outer
boundary of the ring is diffuse along the bar major axis.
On the other hand,
on the bar minor axis the kinematics are strongly affected by the
nuclear ring.
The radial velocity changes from +40 km s\sso\ to -70 km s\sso\ from the inner
to the outer edge of the ring.
The tangential velocity shows an even stronger effect when it quickly
changes from 170 km s\sso\ interior to the ring to 370 km s\sso\ exterior to
the ring.

Because stars, unlike the gas, are not affected by pressure forces, stars
formed in nuclear rings with the initial velocities of the gas will not stay in
the nuclear ring.
In Figure \ref{starorbits} we plot the initial orbits of stars that form 
at either the bar major axis or the bar minor axis. 
In both cases the stellar orbits diverge from the nuclear ring to the outside
after about one-fourth of a rotation.
This divergence occurs over a period of 10 Myr which is well within the lifetime
of an HII region.

\subsection{Comparison to Observations}

The model long slit spectra shown in Figure \ref{allfour}
reveal that the optimum observations to test these
simulations would be to observe gas kinematics along the minor axis of the bar.
It would be best to have the bar minor axis align with either the major
or minor axes of the galaxy to simplify the interpretation.
The strongest signature is in the tangential velocity so the optimum galaxy would
have the minor axis of the bar along the major axis of the galaxy.
The spatial resolution needs to be high enough to prevent beam smearing from destroying
the signal ($\sim$100 pc). 

Our prediction of the young stars migrating to the exterior of the
gas in the nuclear ring is consistent with Hubble Space Telescope (HST) observations
of nuclear rings.
In 
NGC 1512, NGC 5248 \citep{Maoz01},  and NGC 4314  \citep{Benedict02},
the dust ring is interior to the clumps of star formation.
\cite{C92} also reported that the CO ring in NGC 4314 is interior to the radio
continuum ring. 
Although they attributed this to the ring shrinking with time, our models show that
even without shrinking the star formation should be exterior to the gas.
In addition, in a large survey of Seyfert and quiescent galaxies using HST optical
and near infrared observations \cite{Martini02} found seven nuclear ring galaxies,
all of which show the star formation exterior to the dust.

The fact that young stars are seen exterior to the rings of dust and gas in almost
all the observations of nuclear rings is strong evidence against the theory that nuclear
rings are the remnant of a nuclear starburst \citep{KCY93}.
If this theory were correct, then the star formation would be seen interior to the dust and gas.
This is 
because the starburst would have started in the nucleus and proceeded to larger radii.
Therefore,
the unconsumed gas would always be at larger radii that the current star formation.

\subsection{Inner Lindblad Resonances in Bars \label{barilr}}

When Lindblad developed the theory of resonances in a rotating disk he was
working on spiral arm theory \citep{L64}. 
The spiral potential is a small perturbation to the underlying axisymmetric potential
and, thus, the orbits are all nearly circular and, except at resonances,
are not closed in the reference frame of the rotating potential.
Over time this approximation has been expanded to extend to strong bars where:
there exist orbit families that are closed over a large range of radii 
(x$_1$ and x$_2$), the radial
excursions from a circular orbit are a relatively large fraction of the radius, 
and the perturbation to the underlying axisymmetric potential is large.
In this regime the concept of an ILR does not apply.

For spiral arms the orbits are all nearly circular with an
azimuthal frequency of a circular orbit, $\Omega$.
A particle on a circular orbit that undergoes a small perturbation
in a potential that is nearly axisymmetric
will oscillate radially around the circular orbit center with a frequency of $\kappa$.
When $\Omega  \pm \kappa/n$ is equal to the pattern speed of the potential,
 where n is some integer, 
there is a resonance and the orbits
will be closed in the rotating reference frame of the potential.
The resonance is known as an inner Lindblad resonance when 
 $\Omega -\kappa/2$ equals the pattern speed of the potential.
Both $\kappa$ and $\Omega$ vary as a function of the radius based on the
rotation curve.
At the radius where the conditions for the ILR are met, a test particle will exhibit
exactly two radial oscillations per orbit in a reference frame rotating with the potential.
It will, therefore, form a closed orbit in the rotating reference frame and
it will be in the same location relative to the potential after each orbit.
Thus, it will be in an assumed unstable resonance with the potential.

It is important to look at the concept of the ILR relative to the families of
closed bar orbits.
Under the definition of an ILR that the test particle will perform two radial oscillations during
each orbit, both the x$_1$ and x$_2$ orbits are in an ILR-like resonance
because all the orbits in both families perform two radial oscillations per orbit. 
If one looks at the set of x$_1$ orbits that exist in a bar and overlays the 
first order approximation of the ILR radii (Figure \ref{ilrcomp} lower panel),
it is clear that the x$_1$ orbits are unaffected by the ``resonance''.
On the other hand, Figure \ref{ilrcomp} (upper panel) shows that the OEX1R (OILR) is
a rough approximation of the largest extent of x$_2$ orbits along the minor axis of the
bar.

If ILRs do not apply to strong bars, why are they invoked so often to explain
the formation of nuclear rings?
Early work on bar orbits started with the assumption that the bar perturbation was
small relative to the axisymmetric component \citep{CM77}.
In this case, the x$_2$ family of orbits does extend from the IEX2R (IILR) to the OEX2R (OILR).
Over time, this approximation has been extended to stronger bars where it does not apply.
For a weak bar, our results are consistent with \cite{CM77}; 
Figure \ref{hydrox21} shows that the
x$_2$ orbit extents and the IEX2R (IILR) and OEX2R (OILR) match well for weak bars.
However, it is important to remember that the axisymmetric locations of ILRs in strong bars
are not really resonances but merely an approximation of the radii where x$_2$ orbits
become stable. 
From Figures \ref{hydrox22}-\ref{hydrox25} it is also clear that this approximation 
breaks down for strong bars and/or thin bars.
This breakdown in the approximation is well known \citep{vAS82,CG89,A92a}, but many authors 
who do not directly work on bar orbits seem to be unaware of it.
The general misunderstanding arises when authors assume that the bar ILRs are 
resonances in the sense that something unstable happens at these radii.
That is clearly not the case, and, for strong bars, the radii of the predicted
ILRs are meaningless.
Thus, we prefer to refer to the epicyclic predictions of x$_2$ orbital extents rather
than the ILRs.

In summary, for a strong bar the inner Lindblad resonance does not exist
since there is no resonance with the rotating potential at this radius.
In fact, the entire x$_1$ and x$_2$ orbit families are in resonance 
with the potential over their entire extent.
Although the axisymmetric epicyclic predictions of x$_2$ orbit extent (IEX2R and OEX2R)
provide a good estimate of the radial extent of the x$_2$ orbits for
weak bars, this approximation breaks down for strong or thin bars.

\section{Conclusions}

We have shown that nuclear rings result from the interaction of
gas on x$_1$-like and x$_2$-like streamlines.
These rings only occur when a threshold amount of
phase space is occupied by x$_2$ orbits.
The initial radius of the nuclear ring is at the radius of the
x$_2$ orbit with the greatest extent along the bar major axis.
In general, we find that potentials with low quadrupole moments and
strong central concentrations favor nuclear ring formation.

We have shown the radius of nuclear rings decreases with time either
because
the rings accumulate lower specific angular momentum gas or
because of dissipation at the contact point of the bar
dust lane and the nuclear ring.
This decrease of the ring radius with time shows that the 
nuclear ring is not in equilibrium. Over a Hubble time, 
nuclear rings could have disappeared in some galaxies.
This also implies that galaxies with low accretion rates of material
onto their nuclear rings may have longer lived rings.
Also, because there are a range of radii over which a ring could be observed
for a given potential, one cannot directly relate the radius of the
ring to any feature of the rotation curve.

We have also shown that the standard assumption that nuclear rings
are related to inner Lindblad resonances
is based on the approximation that the radial extent of the x$_2$ orbit
family is from the IILR to the OILR.
We have shown that a better name for these radii is the inner epicyclic
x$_2$ radius (IEX2R) and the outer epicyclic x$_2$ radius (OEX2R). 
This epicyclic approximation breaks down for strong or thin bars.
In addition, we have shown that the ILR is not a resonance at all
in barred galaxies.
No resonance occurs at these radii; instead, 
confusion has developed between the orbit family transition that
occurs at these radii in weak bars and a true resonance.

From the gas kinematics in the models, we have predicted the orbits
of stars that form in the nuclear rings.
We have also compared these predictions with Hubble Space
Telescope observations of star forming nuclear rings and shown that
in all cases the star forming regions are exterior to the
nuclear rings of dust and gas as our simulations predict.

\acknowledgements
The authors would like to thank Glen Piner and Jim Stone for providing
the hydrodynamic simulation program that we used in this paper.
The authors also would like to thank the anonymous referee for comments that
improved the paper.

%% ----------------------FIGURES-----------------------------------------
\clearpage
\begin{figure}
\epsscale{0.4}
%plotone{f1.eps}
\caption{x$_1$ and x$_2$ orbits in a barred potential. The bar runs up and
down the page. The x$_1$ orbits provide the primary support for the bar.
The x$_2$ orbits are perpendicular to the bar and do not extend over the
entire bar.}
\label{x1x2orbits}
\end{figure}

\begin{figure}
\epsscale{0.7}
%plotone{f2.eps}
\caption{Gas surface densities from the hydrodynamic simulations for the grid
of potentials. Increased surface density is darker. The bar axis ratio
is 1.5. Each subpanel shows the inner $\pm$5 kpc. The bar is along the y axis.
}
\label{hydro1}
\end{figure}

\begin{figure}
\epsscale{0.88}
%plotone{f3.eps}
\caption{The same as Figure \ref{hydro1} but for a bar axis ratio of 2.0.}
\label{hydro2}
\end{figure}

\begin{figure}
%plotone{f4.eps}
\caption{The same as Figure \ref{hydro1} but for a bar axis ratio of 2.5.}
\label{hydro3}
\end{figure}

\begin{figure}
%plotone{f5.eps}
\caption{The same as Figure \ref{hydro1} but for a  bar axis ratio of 3.0.}
\label{hydro4}
\end{figure}

\begin{figure}
%plotone{f6.eps}
\caption{The same as Figure \ref{hydro1} but for a  bar axis ratio of 3.5.}
\label{hydro5}
\end{figure}

\begin{figure}
%plotone{f7.eps}
\caption{The average density at each radius is plotted in gray scale as a 
function of time. 
The two horizontal
lines are at the radii of the major and minor axis of the smallest x$_2$ orbit.}
\label{denradtime}
\end{figure}

\begin{figure}
%plotone{f8.eps}
\caption{The average density at each radius is plotted in gray scale as a 
function of time. 
For this run the density at radii larger than 1 kpc was lower to 
be less than 1$\times 10^{-5}$M$_{\sun}\ pc^{-2}$ after 300 Myr.
By cutting off the addition of new gas to the nuclear ring 
     the inward migration of the ring is stopped.
}
\label{dentimecutoff}
\end{figure}

\begin{figure}
\epsscale{0.65}
%plotone{f9.eps}
\caption{Locations of x$_2$ orbits (red ellipses) and the inner and
outer epicyclic x$_2$ radii (IILR and OILR) (blue circles) 
overlaid on 
gas surface densities from the hydrodynamic simulations (gray scale).
The outer blue circle is the OEX2R (OILR) and the inner blue circle is the 
IEX2R(IILR).
Each subpanel shows the inner $\pm$1.8 kpc.
The red ellipses show three extreme x$_2$ orbits: the stable orbit with the
largest extent perpendicular to the bar, the stable orbit with the largest
extent parallel to the bar, and the orbit with the smallest 
extent perpendicular to the bar.
In the upper left of each sub-panel is the volume of phase space occupied
by the x$_2$ orbits.
In the lower left of each panel is the simulation run number.
The bar axis ratio is 1.5.}
\label{hydrox21}
\end{figure}

\begin{figure}
\epsscale{0.88}
%plotone{f10.eps}
\caption{The same as Figure \ref{hydrox21} except for a bar axis ratio of 2.0.}
\label{hydrox22}
\end{figure}

\begin{figure}
%plotone{f11.eps}
\caption{The same as Figure \ref{hydrox21} except for a bar axis ratio of 2.5.}
\label{hydrox23}
\end{figure}

\begin{figure}
%plotone{f12.eps}
\caption{The same as Figure \ref{hydrox21} except for a bar axis ratio of 3.0.}
\label{hydrox24}
\end{figure}

\begin{figure}
%plotone{f13.eps}
\caption{The same as Figure \ref{hydrox21} except for a bar axis ratio of 3.5.}
\label{hydrox25}
\end{figure}

\begin{figure}
%plotone{f14.eps}
\caption{The tangential and radial velocities of the gas 
along the bar major and minor axes. The thick line shows the gas
surface density as a function of radius. The thin line shows the velocity 
as a function of radius.}
\label{allfour}
\end{figure}

\begin{figure}
%plotone{f15.eps}
\caption{The orbits that stars forming in the ring would follow.
Each orbit is followed for 10 Myr and shows that the stars migrate to
the exterior of the ring. }
\label{starorbits}
\end{figure}

\begin{figure}
\epsscale{0.17}
%plotone{f16.eps}
\caption{(upper panel) The families of x$_1$ and x$_2$ orbits with the OEX2R (OILR) 
shown as a thick circle.
The OEX2R (OILR) is a reasonable approximation of the radius of the major axis
of the x$_2$ orbit with the largest extent along the bar minor axis.
(lower panel) The x$_1$ orbit family in a barred potential with the OILR 
shown as a thick circle. Note that the OILR has no effect on the x$_1$ orbit
because there is no real resonance at this radius.
}
\label{ilrcomp}
\end{figure}


\clearpage

\begin{thebibliography}{}
\bibitem[Athanassoula(1992a)]{A92a} Athanassoula, E.\ 1992a, \mnras, 259, 328
\bibitem[Athanassoula(1992b)]{A92b} Athanassoula, E.\ 1992b, \mnras, 259, 345
\bibitem[Benedict et al.(2002)]{Benedict02} Benedict, G.~F., 
Howell, D.~A., J{\o}rgensen, I., Kenney, J.~D.~P., \& Smith, B.~J.\ 2002, 
\aj, 123, 1411 
\bibitem[Buta \& Combes(1996)]{BC96} Buta, R.~\& Combes, F.\ 
1996, Fundamentals of Cosmic Physics, 17, 95. 
\bibitem[Byrd et al.(1994)]{B94}Byrd, G., Rautiainen, P., Salo, H., Buta, R., \& Crocker, D. A.
1994, \aj, 108, 476
\bibitem[Combes(1996)]{Combes96} Combes F. 1996, in IAU Colloqium 157,
Barred Galaxies, eds. R. Buta, D. A. Crocker, \& B.G. Elmegreen,
(San Francisco:ASP), 286
\bibitem[Combes \& Gerin(1985)]{CG85}Combes, F. \& Gerin, M. 1985, \aap, 150 327
\bibitem[Combes et al.(1992)]{C92} Combes, F., Gerin, M., Nakai, N., Kawabe, R., \& Shaw, M.~A.\ 1992, \aap, 259, L27
\bibitem[Contopoulos \& Grosbol(1989)]{CG89} Contopoulos, G.~\& Grosbol, P.\ 1989, \aapr, 1, 261
\bibitem[Contopoulos \& Mertzanides(1977)]{CM77}Contopoulos, G. \& Mertzanides, C. 1977,
\aap, 61, 477
\bibitem[Contopoulos \& Papayannopoulos(1980)]{CP80} Contopoulos, G.~\& Papayannopoulos, T.\ 1980, \aap, 92, 33
\bibitem[Garcia-Barreto et al.(1991)]{GB91} Garcia-Barreto, J.~A., Downes, D., 
Combes, F., Gerin, M., Magri, C., Carrasco, L., \& Cruz-Gonzalez, I.\ 1991, \aap, 244, 257
\bibitem[Kenney, Carlstrom, \& Young(1993)]{KCY93} Kenney, J.~D.~P., Carlstrom, 
J.~E., \& Young, J.~S.\ 1993, \apj, 418, 687
\bibitem[Kuzmin(1956)]{K56}Kuzmin, G. 1956, \azh, 33, 27
\bibitem[Lindblad(1964)]{L64}Lindblad, B. 1964, Astrophysica Norvegica, 9, 103
\bibitem[Maoz et al.(2001)]{Maoz01} Maoz, D., Barth, A.~J., Ho, L.~C., Sternberg, A., 
\& Filippenko, A.~V.\ 2001, \aj, 121, 3048
\bibitem[Martini et al.(2002)]{Martini02}Martini, P, Regan, M. W., Mulchaey, J. S.,
\& Pogge, R. W. 2002, \apjs, submitted
\bibitem[Norman, Sellwood, \& Hasan(1996)]{NSH96} Norman, 
C.~A., Sellwood, J.~A., \& Hasan, H.\ 1996, \apj, 462, 114. 
\bibitem[Piner, Stone, \& Teuben(1995)]{PST95} Piner, B.\ G., 
Stone, J.\ M.\ \& Teuben, P.\ J.\ 1995, \apj, 449, 508 
\bibitem[Rautiainen \& Salo(2000)]{RS00}Rautiainen, P. \& Salo, H. 2000, \aap, 362, 465
\bibitem[Regan, Vogel, \& Teuben(1997)]{RVT97} Regan, M.~W., 
Vogel, S.~N., \& Teuben, P.~J.\ 1997, \apjl, 482, L143. 
\bibitem[Regan, Sheth, \& Vogel(1999)]{RSV99} Regan, M.~W., 
Sheth, K., \& Vogel, S.~N.\ 1999, \apj, 526, 97. 
\bibitem[Schwarz(1981)]{S81}Schwarz, M.P., 1981, \apj, 247, 77
\bibitem[Schwarz(1984)]{S84}Schwarz, M.P., 1984, \mnras, 209, 93
\bibitem[Sheth et al.(2000)]{Sheth00} Sheth, K., Regan, M.~W., 
Vogel, S.~N., \& Teuben, P.~J.\ 2000, \apj, 532, 221
\bibitem[Shlosman, Begelman, \& Frank(1990)]{SBF90} Shlosman, I., Begelman, M.~C., \& Frank, J.\ 1990, \nat, 345, 679
\bibitem[Teuben(1995)]{Teuben95} Teuben, P.\ 1995, ASP 
Conf.~Ser.~77: Astronomical Data Analysis Software and Systems IV, 4, 398. 
\bibitem[Toomre(1963)]{T63}Toomre, A. 1963, \apj, 138, 385
\bibitem[van Albada \& Sanders(1982)]{vAS82} van Albada, 
T.~S.~\& Sanders, R.~H.\ 1982, \mnras, 201, 303. 





\end{thebibliography}
\end{document}